\documentstyle[preprint,aps]{revtex}
\tighten
\draft
\widetext
\preprint{YITP-00-74}
\bigskip
\bigskip
\begin{document}
\title{Gravitational Higgs Mechanism}
\medskip
\author{Zurab Kakushadze\footnote{E-mail: 
zurab@insti.physics.sunysb.edu} and Peter Langfelder\footnote{E-mail:
plangfel@insti.physics.sunysb.edu}}
\bigskip
\address{C.N. Yang Institute for Theoretical Physics\\ 
State University of New York, Stony Brook, NY 11794}

\date{November 27, 2000}
\bigskip
\medskip
\maketitle

\begin{abstract} 
{}We discuss the gravitational Higgs mechanism 
in domain wall background solutions that arise in the
theory of 5-dimensional Einstein-Hilbert gravity coupled to 
a scalar field with a non-trivial potential.
The scalar fluctuations in such backgrounds can be completely gauged
away, and so can be the graviphoton fluctuations. On the other hand, we show 
that the graviscalar fluctuations do not have 
normalizable modes. 
As to the 4-dimensional graviton fluctuations,
in the case where the volume of the transverse dimension is finite the massive
modes are plane-wave normalizable, while the zero mode is quadratically 
normalizable. 
We then discuss the coupling of domain wall gravity to localized 
4-dimensional matter. In particular, we point out that this coupling
is consistent only if the matter is conformal. This is different from the
Randall-Sundrum case as there is a discontinuity in the $\delta$-function-like
limit of such a smooth 
domain wall - the latter breaks diffeomorphisms only 
spontaneously, while the Randall-Sundrum brane breaks diffeomorphisms 
explicitly. Finally, at the quantum level
both the domain wall as well as the Randall-Sundrum setups suffer from
inconsistencies in the coupling between gravity and localized matter, 
as well as the fact that gravity is generically expected to be delocalized 
in such backgrounds due to higher curvature terms.  
\end{abstract}

\section{Introduction}

{}In the Brane World scenario the Standard Model gauge and matter fields
are assumed to be localized on  
branes (or an intersection thereof)
embedded in a higher dimensional bulk \cite{early,BK,polchi,witt,lyk,shif,TeV,dienes,3gen,anto,ST,BW,Gog,RS,DGP,DG,alberto}. The volume
of dimensions transverse to the branes is 
automatically finite if these dimensions are compact. On the other hand, 
the volume of the transverse dimensions
can be finite even if the latter are non-compact. In particular, this can be
achieved by using \cite{Gog} warped compactifications \cite{Visser} which
localize gravity on the brane \cite{RS}.

{}A class of examples with localized gravity is given by domain wall solutions
interpolating between two AdS vacua. Such backgrounds spontaneously break
diffeomorphism invariance of the theory, which results in gravitational
Higgs mechanism. One of the aims of this paper is to study the gravitational
Higgs mechanism in some detail. In particular, we compute the spectrum of 
normalizable modes in domain wall background solutions that arise in the
theory of $D$-dimensional  
Einstein-Hilbert gravity coupled to a scalar field with a non-trivial
potential. The scalar fluctuations in such backgrounds can be completely gauged
away, and so can be the graviphoton fluctuations. On the other hand, we show 
that the graviscalar fluctuations do not even have plane-wave 
normalizable modes. As to the $(D-1)$-dimensional graviton fluctuations,
in the case where the domain wall interpolates between two AdS vacua
(so that the volume of the transverse dimension is finite) the massive
modes are plane-wave normalizable, while the zero mode is quadratically 
normalizable. We also discuss the case of domain walls interpolating between
AdS and Minkowski vacua (in this case the volume of the transverse dimension
is infinite) \cite{zuraCBW}, 
where we have the same conclusions as in the previous case
except that the $(D-1)$-dimensional graviton zero mode is no longer
quadratically normalizable but plane-wave normalizable.

{}We then discuss the coupling of domain wall gravity to localized 
$(D-1)$-dimensional matter. In particular, we point out that this coupling
is consistent only if the matter is conformal. This is different from what
happens in the Randall-Sundrum model for the reason that there if a 
discontinuity between the $\delta$-function-like limit of a smooth
domain wall
and the Randall-Sundrum brane - the former breaks diffeomorphisms 
only spontaneously, while the latter breaks some of the diffeomorphisms 
explicitly. We also point out that, in the finite volume cases with
localized gravity, at the quantum level there is an inconsistency 
in the coupling between the domain wall gravity and 
localized matter as the latter generically is no longer conformal. This 
is not unrelated to the fact that at the quantum level gravity is 
generically expected to be delocalized due to higher curvature terms
\cite{COSM,zuraRS,olindo}. 
As to the Randall-Sundrum case, where we also expect that gravity is 
generically expected to be delocalized at the quantum level, an inconsistency
in the coupling between brane world gravity and brane matter is generically
expected to arise due to the fact that in this case
the graviscalar does not decouple in the ultra-violet \cite{zuraRS}.

{}Finally, we point out that the aforementioned difficulties do not arise
in the recent proposal of \cite{alberto}, where we have completely localized
gravity on a solitonic brane. In particular, in this case there are no 
propagating degrees of freedom in the bulk, and no inconsistency related to
higher curvature terms or graviscalar coupling is expected 
to arise at the quantum level.    

\section{Gravity in Domain Wall Backgrounds}

{}Consider a single real scalar field $\phi$
coupled to gravity with the following action\footnote{Here we focus on
the case with one scalar field for the sake of simplicity. In particular,
in this case we can absorb a (non-singular) metric $Z(\phi)$ in
the $(\nabla\phi)^2$ term by a non-linear field redefinition. This cannot
generically be done in the case of multiple scalar fields $\phi^i$, where
one must therefore also consider the metric $Z_{ij}(\phi)$.}:
\begin{equation}
 S=M_P^{D-2}\int d^Dx \sqrt{-G}\left[R-
 {4\over D-2}(\nabla \phi)^2 -V(\phi)\right]~,
 \label{actionphi}
\end{equation} 
where $M_P$ is the $D$-dimensional (reduced) Planck scale,
and $V(\phi)$ is the scalar potential for $\phi$. 
The equations of motion read:
\begin{eqnarray}
 && {8\over{D-2}}\nabla^2\phi=V_\phi~,\\
 \label{einstein1}
 && R_{MN}-{1\over 2}G_{MN} R
 ={4\over {D-2}}\left[\nabla_M\phi\nabla_N\phi
 -{1\over 2}G_{MN}(\nabla \phi)^2\right]-{1\over 2}G_{MN} V~.
\end{eqnarray}
The subscript $\phi$ in $V_\phi$ denotes derivative w.r.t. $\phi$.

{}In the following we will be interested in solutions to the above equations
of motion where the metric has the following warped \cite{Visser} form
\begin{equation}\label{warpedy}
 ds^2=\exp(2A)\eta_{MN}dx^M dx^N~,
\end{equation}
where $\eta_{MN}$ is the flat $D$-dimensional Minkowski metric,
and the warp factor $A$ 
and the scalar field $\phi$ are non-trivial functions
of $z\equiv x^D$ but are independent of the other $(D-1)$ coordinates
$x^\mu$.
With this Ansatz we have the following equations of motion   
for $\phi$ and $A$ (prime denotes derivative w.r.t. $z$): 
\begin{eqnarray}
 \label{phi''1}
 &&{8\over{D-2}}\left[\phi^{\prime\prime}+(D-2)A^\prime
 \phi^\prime\right]-V_\phi\exp(2A)=0~,\\
 \label{phi'A'1}
 &&(D-1)(D-2)(A^\prime)^2
 -{4\over D-2}(\phi^\prime)^2+V\exp(2A)=0~,\\
 \label{A''1}
 &&(D-2)\left[A^{\prime\prime}-(A^\prime)^2\right]
 +{4\over D-2}(\phi^\prime)^2=0~.
\end{eqnarray}
We can rewrite these equations in terms of
the following first order equations
\begin{eqnarray}\label{BPS1}
 &&\phi^\prime=\alpha W_\phi\exp(A)~,\\
 \label{BPS2}
 &&A^\prime=\beta W\exp(A)~,
\end{eqnarray}
where 
\begin{eqnarray}
 &&\alpha\equiv\sigma {\sqrt{D-2}\over 2}~,\\
 &&\beta\equiv-\sigma {2\over (D-2)^{3/2}}~,
\end{eqnarray}
and $\sigma=\pm 1$. Moreover, the scalar potential $V$ is related
to the function $W=W(\phi)$ via
\begin{equation}
\label{potential}
 V=W_\phi^2-\gamma^2 W^2~,
\end{equation}
where
\begin{equation}
 \gamma^2\equiv {4(D-1)\over (D-2)^2}~.
\end{equation}
In the supersymmetric context the function $W(\phi)$ is interpreted as the
superpotential, while the equations (\ref{BPS1}) and (\ref{BPS2}) are the
BPS equations, which imply that the domain wall breaks 1/2 of the original
supersymmetries.

\begin{center}
 {\em A Simple Example}
\end{center}

{}Let us give a simple example of a domain wall solution of the above type.
Thus, let
\begin{equation}
 W=\xi\left[\zeta\phi-{1\over 3}\zeta^3\phi^3\right]~,
\end{equation}
where $\xi$ and $\zeta$ are parameters. 
The domain wall solution is then given by:
\begin{eqnarray}\label{smooth1}
 &&\phi(y)={1\over\zeta}\tanh\left[\alpha\xi\zeta^2
 (y-y_0)\right]~,\\
 \label{smooth2}
 &&A(y)={2\beta\over{3\alpha\zeta^2}}\left\{\ln\left(
 \cosh\left[\alpha\xi\zeta^2 (y-y_0)\right]\right)-{1\over 4}{1\over
 \cosh^2\left[\alpha\xi\zeta^2 (y-y_0)\right]}\right\}+A_0~,
\end{eqnarray}
where $y_0$ and $A_0$ are integration constants, which we will set to
zero in the following. Then the point $y=0$ corresponds to the ``center''
of the domain wall, and at this point the warp factor vanishes: $A(0)=0$.
For convenience reasons here instead of the $(x^\mu,z)$ coordinate system 
we are using the $(x^\mu,y)$ coordinate system, where
\begin{equation}\label{map}
 dy=\exp(A) dz~.
\end{equation}
In the following we will set the integration constant in the solution $y=y(z)$
of (\ref{map}) so that the $z=0$ point corresponds to the $y=0$ point. 

{}Note that in the above solution the volume of the $y$ direction, which is
given by
\begin{equation}\label{volume}
 v=\int dy \exp[(D-1)A]~,
\end{equation}
is finite. This implies that gravity is localized on the domain wall.
In the following we will be interested in precisely such domain walls.

\subsection{Normalizable Modes}

{}Let us now study gravity in the above type of smooth domain wall 
backgrounds. 
In particular, here we would like to compute the spectrum of normalizable
modes. Thus, let us consider small fluctuations around the domain wall solution
(\ref{smooth1}) and (\ref{smooth2}): 
\begin{equation}\label{fluctu}
 G_{MN}=\exp(2A)\left[\eta_{MN}+{\widetilde h}_{MN}\right]~,
\end{equation}
where for convenience reasons we have chosen to work with 
${\widetilde h}_{MN}$
instead of metric fluctuations $h_{MN}=\exp(2A){\widetilde h}_{MN}$.
Also, let $\varphi$ be the fluctuation of the scalar field around the
background $\phi=\phi(z)$.

{}In terms of ${\widetilde h}_{MN}$ the full
$D$-dimensional diffeomorphisms (corresponding to $x^M\rightarrow x^M-\xi^M$)
\begin{equation}
 \delta h_{MN}=\nabla_M\xi_N+\nabla_N\xi_M
\end{equation}
are given by the following gauge 
transformations (here we use $\xi_M\equiv \exp(2A){\widetilde \xi}_M$):
\begin{equation}\label{gauge}
 \delta{\widetilde h}_{MN}=\partial_M {\widetilde\xi}_N+
 \partial_N{\widetilde\xi}_M+2A^\prime\eta_{MN} \omega~, 
\end{equation} 
where $\omega\equiv {\widetilde \xi}_D$. As to the scalar field $\varphi$,
we have:
\begin{equation}
 \delta\varphi=\phi^\prime\omega~.
\end{equation}
Since the domain wall solution does not break diffeomorphisms explicitly
but spontaneously,
the linearized equations of motion are invariant under the full
$D$-dimensional diffeomorphisms.

{}In the following we will use the following notations for the component
fields:
\begin{equation}
 H_{\mu\nu}\equiv{\widetilde h}_{\mu\nu}~,~~~A_\mu\equiv
 {\widetilde h}_{\mu D}~,~~~\rho\equiv{\widetilde h}_{DD}~.
\end{equation}
In terms of the component fields $H_{\mu\nu}$, $A_\mu$ and $\rho$, the
full $D$-dimensional diffeomorphisms read:
\begin{eqnarray}\label{diff1}
 &&\delta H_{\mu\nu}=\partial_\mu{\widetilde\xi}_\nu+\partial_\nu{\widetilde
 \xi}_\mu+2\eta_{\mu\nu}A^\prime\omega~,\\
 \label{diff2}
 &&\delta A_\mu={\widetilde\xi_\mu}^\prime +\partial_\mu \omega~,\\
 \label{diff3}
 &&\delta\rho=2\omega^\prime+2A^\prime\omega~,\\
 \label{diff4}
 &&\delta\varphi=\phi^\prime\omega~.
\end{eqnarray}
In the following we will also use the notation $H\equiv {H_\mu}^\mu$.

{}The linearized equations of motion read:
\begin{eqnarray}\label{EOM1oGBsm}
 &&
 \partial_\sigma\partial^\sigma 
 H_{\mu\nu} +\partial_\mu\partial_\nu
 H-\partial_\mu \partial^\sigma H_{\sigma\nu}-
 \partial_\nu \partial^\sigma H_{\sigma\mu}-\eta_{\mu\nu}
 \left[\partial_\sigma\partial^\sigma H-\partial^\sigma\partial^\rho
 H_{\sigma\rho}\right]+\nonumber\\
 &&H_{\mu\nu}^{\prime\prime}-\eta_{\mu\nu}H^{\prime\prime}+
 (D-2)A^\prime\left[H_{\mu\nu}^\prime-\eta_{\mu\nu}H^\prime\right]-
 \nonumber\\
 &&\left\{\partial_\mu A_\nu^\prime + \partial_\nu A_\mu^\prime -
 2\eta_{\mu\nu}\partial^\sigma A_\sigma^\prime+ (D-2)A^\prime
 \left[\partial_\mu A_\nu +\partial_\nu A_\mu
 -2\eta_{\mu\nu}\partial^\sigma
 A_\sigma\right]\right\} +\nonumber\\
 &&\partial_\mu\partial_\nu\rho-\eta_{\mu\nu}
 \partial_\sigma\partial^\sigma 
 \rho+\eta_{\mu\nu}\left[(D-2)A^\prime\rho^\prime
 -V\exp(2A) \rho\right]=\nonumber\\
 &&{8\over{D-2}}\eta_{\mu\nu}\phi^\prime\varphi^\prime+\eta_{\mu\nu}
 \varphi V_\phi\exp(2A)~,\\ 
 \label{EOM2oGBsm} 
 &&
 \left[\partial^\mu H_{\mu\nu}-\partial_\nu H\right]^\prime 
 -\partial^\mu F_{\mu\nu}+
 (D-2)A^\prime \partial_\nu\rho={8\over{D-2}}\phi^\prime\partial_\nu
 \varphi~,\\
 \label{EOM3oGBsm}
 &&-\left[\partial^\mu\partial^\nu H_{\mu\nu}-\partial^\mu\partial_\mu H
 \right] +(D-2) A^\prime \left[H^\prime-2\partial^\sigma A_\sigma\right]
 +V\exp(2A)\rho =\nonumber\\
 &&{8\over{D-2}}\phi^\prime\varphi^\prime-\varphi V_\phi\exp(2A)~,\\
 \label{EOM4oGBsm}
 &&\partial_\mu\partial^\mu\varphi+\varphi^{\prime\prime}+(D-2)A^\prime
 \varphi^\prime-{{D-2}\over 8}\varphi V_{\phi\phi}\exp(2A)-\nonumber\\
 &&{1\over 2}\phi^\prime\left[2\partial^\mu A_\mu+\rho^\prime-
 H^\prime\right]-{{D-2}\over 8}\rho V_\phi\exp(2A)=0~,
\end{eqnarray}
where $F_{\mu\nu}\equiv\partial_\mu A_\nu-\partial_\nu A_\mu$ is the 
$U(1)$ field strength for the graviphoton.

{}Next, let us note that the field $\varphi$ can be completely  
eliminated from the
above equations of motion \cite{COSM,zuraCBW}. 
Indeed, this is achieved via diffeomorphisms with
\begin{equation}
 \omega=-\varphi/\phi^\prime~.
\end{equation}  
That is, $\varphi$ is {\em not}
a propagating degree of freedom in this gauge \cite{COSM,zuraCBW}. This is an
important point, which implies that not only the $\varphi$ zero mode but the
entire field $\varphi$ is ``eaten'' in the gravitational Higgs mechanism. 

{}Note that setting $\varphi$ to zero uses up some
diffeomorphisms, but the residual diffeomorphisms are sufficient to also gauge
$A_\mu$ away. Indeed, after we remove $\varphi$ from the equations of motion,
we can use the diffeomorphisms with $\omega\equiv 0$ but non-trivial
${\widetilde\xi}_\mu$ to set $A_\mu$ to zero without otherwise changing the
form of the equations of motion. We, therefore, obtain:  
\begin{eqnarray}\label{EOM1oGBsmX}
 &&\partial_\sigma\partial^\sigma 
 H_{\mu\nu} +\partial_\mu\partial_\nu
 H-\partial_\mu \partial^\sigma H_{\sigma\nu}-
 \partial_\nu \partial^\sigma H_{\sigma\mu}-\eta_{\mu\nu}
 \left[\partial_\sigma\partial^\sigma H-\partial^\sigma\partial^\rho
 H_{\sigma\rho}\right]+\nonumber\\
 &&H_{\mu\nu}^{\prime\prime}-\eta_{\mu\nu}H^{\prime\prime}+
 (D-2)A^\prime\left[H_{\mu\nu}^\prime-\eta_{\mu\nu}H^\prime\right]+
 \nonumber\\
 &&\partial_\mu\partial_\nu\rho-\eta_{\mu\nu}
 \partial_\sigma\partial^\sigma 
 \rho+\eta_{\mu\nu}\left[(D-2)A^\prime\rho^\prime-V\exp(2A)\rho\right]
 =0~,\\ 
 \label{EOM2oGBsmX} 
 &&\left[\partial^\mu H_{\mu\nu}-\partial_\nu H\right]^\prime 
 +(D-2)A^\prime \partial_\nu\rho=0~,\\
 \label{EOM3oGBsmX}
 &&-\left[\partial^\mu\partial^\nu H_{\mu\nu}-\partial^\mu\partial_\mu H
 \right] +(D-2) A^\prime H^\prime+V\exp(2A)\rho=0~,\\
 \label{EOM4oGBsmX}
 &&\phi^\prime\left[\rho^\prime-
 H^\prime\right]+{{D-2}\over 4}\rho V_\phi\exp(2A)=0~.
\end{eqnarray}
Here we note that the graviscalar component cannot be gauged away 
after we perform the above gauge fixing.

{}Here we note that not all of the above equations are independent.
Thus, differentiating (\ref{EOM1oGBsmX}) with $\partial^\mu$, we obtain  
an equation which is identically satisfied once we take into account
(\ref{EOM2oGBsmX}) as well as the on-shell expressions for $A$ and $\phi$.
Also, if we take the trace of (\ref{EOM1oGBsmX}), then we obtain an equation
which is identically satisfied once we take into account (\ref{EOM2oGBsmX}),
(\ref{EOM3oGBsmX}) and (\ref{EOM4oGBsmX}) as well as the on-shell 
expressions for $A$ and $\phi$. 
This, as usual, is a consequence of Bianchi identities.

{}Now we are ready to discuss normalizable modes in the above domain wall
background. Let us first consider the normalizable modes for the graviscalar
$\rho$. Thus, we can eliminate $H_{\mu\nu}$ from (\ref{EOM2oGBsmX}),
(\ref{EOM3oGBsmX}) and (\ref{EOM4oGBsmX}), which gives us the following
second order equation for $\rho$:
\begin{equation}\label{moderho}
 \rho^{\prime\prime}+\psi A^\prime\rho^\prime+\partial^\mu\partial_\mu\rho
 +F\rho=0~,
\end{equation}  
where
\begin{equation}
 \psi(z)\equiv D+{2\alpha\over\beta} {W_{\phi\phi}\over W}~,
\end{equation}
and
\begin{equation}
 F(z)\equiv 2\exp(2A)\left[(D-1)\beta^2 W^2-W_\phi^2+
 2\alpha\beta W W_{\phi\phi} +\alpha^2 W_\phi W_{\phi\phi\phi}
 \right]~.
\end{equation}
Let us now assume that $\rho$ satisfies the $(D-1)$-dimensional Klein-Gordon 
equation
\begin{equation}
 \partial^\mu\partial_\mu\rho=m^2\rho~.
\end{equation}
In the following we will assume that $m^2\geq 0$. As to the $m^2<0$ modes,
they cannot be normalizable - indeed, the domain wall is a kink-like 
object, and is therefore stable, so no tachyonic modes are normalizable.

{}We need to understand the asymptotic behavior of $\rho$ at large $z$.
To do this, let us first note that at large $z$ the function $\psi(z)$
goes to a constant asymptotic value:
\begin{equation}
 \psi(z\rightarrow\pm\infty)\equiv \psi_0~.
\end{equation}
Here for simplicity we are assuming that $W(-\phi)=-W(\phi)$, so that the 
asymptotic values of $\psi(z)$ at $z\rightarrow\pm\infty$ are the same.
Also, note that 
\begin{equation}\label{rhoD}
 \psi_0>D~.
\end{equation}
Thus, for instance, in the example given by (\ref{smooth1}) and (\ref{smooth2})
we have 
\begin{equation}
 \psi_0=D+{3\over 2}(D-2)^2\zeta^2~.
\end{equation}
In the following, since we are interested in the asymptotic behavior of $\rho$ 
at large $z$, we will replace $\psi(z)$ in (\ref{moderho}) by $\psi_0$.

{}To proceed further, it is convenient to rescale $\rho$ as follows:
\begin{equation}
 \rho\equiv{\widetilde \rho}\exp\left[-{1\over 2}\psi_0 A\right]~.
\end{equation}
At large $z$ the equation (\ref{moderho}) then reads:
\begin{equation}
 {\widetilde \rho}^{\prime\prime}+\left[m^2+F-{1\over 2}\psi_0
 A^{\prime\prime}-{1\over 4}\psi_0^2(A^\prime)^2\right]{\widetilde \rho}=0~.
\end{equation}
Note that at large $z$ the functions $F$, $A^{\prime\prime}$ and
$(A^\prime)^2$ go to zero as $\sim 1/z^2$. We therefore have the following
leading behavior for ${\widetilde \rho}$ at large $z$:
\begin{equation}
 {\widetilde \rho}(z)=C_1 \cos(mz)+C_2 \sin(mz)~,
\end{equation}
where $C_1,C_2$ are some constant coefficients.

{}Next, note that the norm for the graviscalar is given by
\begin{equation}
 ||\rho||^2\propto\int dz~\exp(DA)\rho^2~,
\end{equation}
where
the measure $\exp(DA)$ comes from $\sqrt{-G}$. In terms of ${\widetilde
\rho}$ we have
\begin{equation}
 ||\rho||^2\propto\int dz~\exp\left[(D-\psi_0)A\right]
 {\widetilde \rho}^2~.
\end{equation}
Since $A$ goes to $-\infty$ at large $z$, we conclude that, due to 
(\ref{rhoD}), none of the $m^2>0$ modes
are even plane-wave normalizable. Moreover, since the 
function $F$ in (\ref{moderho}) is non-trivial, we do not have a quadratically
normalizable zero mode either. Thus, we conclude that $\rho$ is {\em not} a
propagating degree of freedom in the above background, and should be set to
zero.

{}Next, let us turn to the normalizable modes for the graviton 
$H_{\mu\nu}$. From (\ref{EOM2oGBsmX}),
(\ref{EOM3oGBsmX}) and (\ref{EOM4oGBsmX}) it follows that, since $\rho
\equiv 0$, we have
\begin{equation}\label{TTz}
 \partial^\mu H_{\mu\nu}^\prime=H^\prime=0~.
\end{equation}
This then implies that we can use the residual $(D-1)$-dimensional
diffeomorphisms (for which $\omega\equiv 0$, and ${\widetilde\xi}_\mu$ are 
independent of $z$) to bring $H_{\mu\nu}$ into the transverse-traceless
form:
\begin{equation}
 \partial^\mu H_{\mu\nu}=H=0~.
\end{equation}
It then follows from (\ref{EOM1oGBsmX}) that for the modes of the form 
\begin{equation}
 H_{\mu\nu}=\xi_{\mu\nu}(x^\rho) \Sigma(z)~,
\end{equation}
where
\begin{equation}
 \partial^\sigma\partial_\sigma \xi_{\mu\nu}=m^2\xi_{\mu\nu}~,
\end{equation}
the $z$-dependent part of $H_{\mu\nu}$ satisfies the following equation:
\begin{equation}\label{modeH}
 \Sigma^{\prime\prime} +(D-2)A^\prime\Sigma^\prime +m^2 
 \Sigma=0~,
\end{equation}
Let us rescale $\Sigma$ as follows:
\begin{equation}
 \Sigma\equiv{\widetilde\Sigma}\exp\left[-{1\over 2}(D-2)A\right]~.
\end{equation}
The equation (\ref{modeH}) now reads:
\begin{equation}
 {\widetilde\Sigma}^{\prime\prime}+\left[m^2 -{1\over 2}(D-2) 
 A^{\prime\prime}-{1\over 4}(D-2)^2(A^\prime)^2\right]{\widetilde\Sigma}=0~.
\end{equation}
At large $z$ we therefore have:
\begin{equation}
 {\widetilde \Sigma}(z)=D_1 \cos\left(m z\right) +D_2 
 \sin\left(m z\right)~,
\end{equation}
where $D_1,D_2$ are some constant coefficients. 

{}Next, note that the norm for the graviton is given by
\begin{equation}
 ||H_{\mu\nu}||^2\propto\int dz~\exp[(D-2)A]\Sigma^2~,
\end{equation}
where, unlike the graviscalar case,
the measure $\exp[(D-2)A]$ comes from $\sqrt{-G}R$. In terms of ${\widetilde
\Sigma}$ we have
\begin{equation}
 ||H_{\mu\nu}||^2
 \propto\int dz~
 {\widetilde \Sigma}^2~.
\end{equation}
Thus, we see that the $m^2>0$ modes of $H_{\mu\nu}$ are 
plane-wave normalizable. Moreover, we also
have a quadratically normalizable zero mode for $H_{\mu\nu}$.
This zero mode is given by $\Sigma^\prime=0$. 

{}Thus, as we see, in smooth domain wall backgrounds 
of the aforementioned type only the $H_{\mu\nu}$ components correspond to
propagating degrees of freedom, while others either can be gauged away or
do not have normalizable modes. This is a result of the gravitational Higgs
mechanism. 

\begin{center}
 {\em Infinite Volume Cases}
\end{center}

{}We would like to end this subsection with the following remark. Above we
considered domain walls interpolating between two AdS vacua. Here we note that
we can also have domain walls interpolating between AdS and Minkowski vacua
\cite{zuraCBW}. 
Let us consider a simple example of such a domain wall. Thus, let
\begin{equation}
 W=\xi\left[\zeta\phi-{1\over 3}\zeta^3\phi^3-{2\over 3}\right]~,
\end{equation}
where $\xi$ and $\zeta$ are parameters. 
The domain wall solution is then given by:
\begin{eqnarray}\label{smooth1X}
 \phi(y)=&&{1\over\zeta}\tanh\left[\alpha\xi\zeta^2
 (y-y_0)\right]~,\\
 \label{smooth2X}
 A(y)=&&{2\beta\over{3\alpha\zeta^2}}\left\{\ln\left(
 \cosh\left[\alpha\xi\zeta^2 (y-y_0)\right]\right)-{1\over 4}{1\over
 \cosh^2\left[\alpha\xi\zeta^2 (y-y_0)\right]}\right\}-\nonumber\\
 &&{2\over 3}\beta\xi (y-y_0)+A_0~,
\end{eqnarray}
where, as before, $y_0$ and $A_0$ are integration constants.

{}Note that in the above solution the volume of the $y$ direction, which is
given by (\ref{volume}),
is {\em infinite}. 
This implies that gravity is {\em not} localized on the domain wall.
The above computations for the spectrum of normalizable modes can be
straightforwardly applied to this case as well. In particular, it is not
difficult to show that, as before, 
the only normalizable modes are those corresponding to
the $(D-1)$-dimensional graviton $H_{\mu\nu}$. The difference, however, is that
all of the modes $m^2\geq 0$ are only plane-wave normalizable, that is, we do 
not have a quadratically normalizable zero mode in this case.

\subsection{Comparison with the Global Case}

{}For comparative purposes 
we would like to end this section by briefly discussing
what happens if we turn off gravity. In this case the relevant action is
\begin{equation}
 {\cal S}=\int d^D x \left[-{4\over {D-2}}(\partial\phi)^2-{\cal V}(\phi)
 \right]~,
\end{equation}
where the potential ${\cal V}$ is given by
\begin{equation}
 {\cal V}=W_\phi^2~.
\end{equation}
Let us focus on backgrounds where $\phi$ depends non-trivially on 
$x^D\equiv y$ but is independent of the other $(D-1)$ coordinates $x^\mu$. 
The equation of motion for $\phi$ can then be written as
\begin{equation}
 \phi_y=\alpha W_\phi~.
\end{equation} 
As before, let
\begin{equation}
 W=\xi\left[\zeta\phi-{1\over 3}\zeta^3\phi^3\right]~,
\end{equation}
where $\xi$ and $\zeta$ are parameters. 
The domain wall solution is then given by:
\begin{equation}\label{smoothglobal}
 \phi(y)={1\over\zeta}\tanh\left[\alpha\xi\zeta^2
 (y-y_0)\right]~,
\end{equation}
where $y_0$ corresponds to the ``center'' of the domain wall, which, 
in this case, is a kink. In the following we will set $y_0=0$.

{}Let us now discuss the spectrum of normalizable modes. The equation of motion
for small fluctuations $\varphi$ around the background is given by
\begin{equation}
 \varphi_{yy}+\partial^\mu\partial_\mu\varphi-\alpha^2
 \left[W_{\phi\phi}^2+W_{\phi}W_{\phi\phi\phi}\right]\varphi=0~.
\end{equation}
Let us assume that $\varphi$ satisfies the $(D-1)$-dimensional Klein-Gordon 
equation
\begin{equation}
 \partial^\mu\partial_\mu\varphi=m^2\varphi~.
\end{equation}
In the following we will assume that $m^2\geq 0$. As to the $m^2<0$ modes, 
they are not normalizable as the above kink background is stable.

{}In the background (\ref{smoothglobal}) we have:
\begin{equation}
 \varphi_{yy}+\left\{m^2-4m_*^2\left[1-{3\over 2
 \cosh^2\left(m_* y\right)}\right]\right\}\varphi=0~,
\end{equation}
where 
\begin{equation}
 m_*\equiv |\alpha\xi|\zeta^2~.
\end{equation}
The spectrum of normalizable modes is then as follows
\cite{early}. There is a quadratically
normalizable zero mode given by (note that in this setup the measure in the
norm of the $\varphi$ field is trivial) 
\begin{equation}
 \varphi(x^\mu,y)={\chi(x^\mu)\over\cosh^2\left(m_*y
 \right)}~,
\end{equation}
where $\chi(x^\mu)$ satisfies the massless $(D-1)$-dimensional Klein-Gordon
equation. The modes with masses $0<m\leq 2m_*$ are not normalizable
except for an isolated mode with $m^2=3 m_*^2$, which is quadratically
normalizable. Finally,
the modes with masses $m>2m_*$ are plane-wave normalizable. Thus, we have a
mass gap in this model. The zero mode is the Goldstone mode of 
the broken translational invariance in the $y$ direction. Upon gauging the
diffeomorphisms, that is, once we include gravity, we expect that this mode is
eaten in the corresponding gravitational Higgs 
mechanism. As we saw in the previous
subsection, this is indeed the case. Note, however, that in the gravitational 
Higgs mechanism not only the zero mode but {\em all} the other $\varphi$ modes
are eliminated 
as well including the aforementioned isolated massive quadratically
normalizable mode with $m^2=3m_*^2$.

\section{Coupling to Localized Matter}

{}In this section we would like to study how gravity in the domain wall 
background couples to localized matter. To do this, let us introduce
a {\em probe} $\delta$-function-like codimension one
brane with matter localized on it. We will refer to the hypersurface 
corresponding to this brane as $\Sigma$, and we will denote its location 
in the $z$ direction via $z_0$. 
For this probe brane not to affect the domain wall background, it must be
{\em tensionless}, and its coupling to the scalar field $\phi$ must vanish as
well.

{}The term describing the interaction of brane matter with bulk fields is 
given by 
\begin{equation}
 S_{\rm {\small int}}=\int_\Sigma d^{D-1} x \left[{1\over 2} T_{\mu\nu}
 H^{\mu\nu}+{8\over{D-2}}\Theta\varphi\right]~,
\end{equation} 
where $T_{\mu\nu}$ is the (properly normalized) energy-momentum tensor,
and $\Theta$ is the coupling of $\varphi$ to the brane matter. The invariance
under the diffeomorphisms (\ref{diff1}) and (\ref{diff4}) implies that
the energy-momentum tensor is conserved
\begin{equation}
 \partial^\mu T_{\mu\nu}=0~,
\end{equation}
while the trace of the energy momentum tensor and the coupling to the scalar
field must satisfy the following condition (note that $\phi^\prime$ does not
vanish anywhere):
\begin{equation}\label{condition0}
 \Theta=-{{D-2}\over 8}{A^\prime(z_0)\over\phi^\prime(z_0)}T~,
\end{equation} 
where $T\equiv {T_\mu}^\mu$. 

{}To proceed further, we need equations of motion for small perturbations in 
the presence of the above matter sources. As before, in these equations we can
gauge $A_\mu$ and $\varphi$ away. Then the resulting equations of motion read:
\begin{eqnarray}\label{EOM1oGBsmXYY}
 &&\partial_\sigma\partial^\sigma 
 H_{\mu\nu} +\partial_\mu\partial_\nu
 H-\partial_\mu \partial^\sigma H_{\sigma\nu}-
 \partial_\nu \partial^\sigma H_{\sigma\mu}-\eta_{\mu\nu}
 \left[\partial_\sigma\partial^\sigma H-\partial^\sigma\partial^\rho
 H_{\sigma\rho}\right]+\nonumber\\
 &&H_{\mu\nu}^{\prime\prime}-\eta_{\mu\nu}H^{\prime\prime}+
 (D-2)A^\prime\left[H_{\mu\nu}^\prime-\eta_{\mu\nu}H^\prime\right]+
 \nonumber\\
 &&\partial_\mu\partial_\nu\rho-\eta_{\mu\nu}
 \partial_\sigma\partial^\sigma 
 \rho+\eta_{\mu\nu}\left[(D-2)A^\prime\rho^\prime-V\exp(2A)\rho\right]
 =-M_P^{2-D} T_{\mu\nu}\delta(z-z_0)~,\\ 
 \label{EOM2oGBsmXYY} 
 &&\left[\partial^\mu H_{\mu\nu}-\partial_\nu H\right]^\prime 
 +(D-2)A^\prime \partial_\nu\rho=0~,\\
 \label{EOM3oGBsmXYY}
 &&-\left[\partial^\mu\partial^\nu H_{\mu\nu}-\partial^\mu\partial_\mu H
 \right] +(D-2) A^\prime H^\prime+V\exp(2A)\rho=0~,\\
 \label{EOM4oGBsmXYY}
 &&\phi^\prime\left[\rho^\prime-
 H^\prime\right]+{{D-2}\over 4}\rho V_\phi\exp(2A)=2M_P^{2-D}\Theta
 \delta(z-z_0)~.
\end{eqnarray}
Note that 
the l.h.s. of the last equation contains no terms with the second
derivative w.r.t. $z$, while for non-vanishing scalar coupling $\Theta$
the r.h.s. contains a $\delta$-function source
term. This implies that this equation does not have a consistent solution
unless we require that
\begin{equation}
 \Theta=0~.
\end{equation}  
This together with (\ref{condition0}) implies that, unless $A^\prime(z_0)=0$,
we must have $T=0$. To avoid this restriction on the trace of the 
energy-momentum tensor, we must place the brane at the ``center'' of the
domain wall (that is, we must set $z_0=0$), where we have 
$A^\prime(0)=0$. (Note that if the domain wall interpolates
between AdS and Minkowski vacua
$A^\prime$ is non-vanishing everywhere, and
the consistency requires that the brane matter be conformal 
\cite{zuraCBW}.)     

{}Note that, with $\Theta=0$, as before
(\ref{EOM2oGBsmXYY}), (\ref{EOM3oGBsmXYY}) and
(\ref{EOM4oGBsmXYY}) imply that $\rho$ must be set to zero. Moreover, we still
have (\ref{TTz}). It is then not difficult to see that (\ref{EOM1oGBsmXYY})
can be satisfied if and only if
\begin{equation}
 T=0~,
\end{equation}
that is, the localized matter must be {\em conformal}. If this condition
is satisfied, then the solution for the graviton field $H_{\mu\nu}$ is given
by
\begin{equation}
 H_{\mu\nu}(p,z)=M_P^{2-D}\Omega(p,z) T_{\mu\nu}~,
\end{equation}
where $\Omega(p,z)$ is the solution to the following equation
\begin{equation}
 \Omega^{\prime\prime}(p,z)+(D-2)A^\prime\Omega^\prime(p,z)-p^2\Omega(p,z)=-
 \delta(z)
\end{equation}
subject to the boundary conditions (for $p^2\equiv p_\mu p^\mu>0$)
\begin{equation}
 \Omega(p,z\rightarrow\pm \infty)=0~.
\end{equation}
Here we have Fourier transformed the $(D-1)$ coordinates $x^\mu$ ($p^\mu$
are the corresponding momenta), and Wick rotated to the Euclidean space (where
the propagator is unique). 
The above solution describes a gravitational field of conformal matter
localized on the brane.

\subsection{The $\delta$-function-like Limit and a Discontinuity}

{}The above result, that localized matter cannot be consistently coupled to
domain wall gravity unless the former is conformal, might at first appear
surprising. In particular, naively it might seem that in the thin wall 
limit, where the domain wall becomes $\delta$-function like, one should 
reproduce the setup of \cite{RS}, where it appears that 
non-conformal matter can, at least at the classical level, be coupled to bulk
gravity. This, however, is not so for a simple reason which we would like to
discuss next.

{}To begin with, let us note that if we take the limit $\zeta\rightarrow\infty$
in the domain wall solution (\ref{smooth1}) and (\ref{smooth2}), we obtain
a $\delta$-function-like brane solution with vanishing scalar field and the
warp factor
\begin{equation}\label{limit}
 A(y)=-{|y|\over \Delta}~,
\end{equation}
where 
\begin{equation}
 \Delta\equiv {3\over 2\beta\xi}~.
\end{equation}
This warp factor is of the same form as in the Randall-Sundrum model
\cite{RS}, where we have a codimension one
brane with tension $f>0$ embedded in the bulk with 
constant vacuum energy density $\Lambda<0$:
\begin{equation}\label{RSaction}
 S=-f\int_{\rm{\small brane}} d^{D-1}x\sqrt{-{\widehat G}}+
 M_P^{D-2}
 \int d^D x \sqrt{-G} \left[R-\Lambda\right]~,
\end{equation}
where
\begin{equation}
 {\widehat G}_{\mu\nu}\equiv{\delta_\mu}^M {\delta_\nu}^N G_{MN}
 \Big|_{z=0}~.
\end{equation}
Here for definiteness we are assuming that the brane is located at $z=0$.
With the appropriately fine-tuned brane tension $f$ and bulk vacuum energy
density $\Lambda$ in this model we then have a solution with
precisely the warp factor of the form (\ref{limit}) and $(D-1)$-dimensional
Poincar{\'e} invariance on the brane.

{}There is, however, a crucial difference between the above two setups.
The smooth domain wall solution, even in the aforementioned limit, breaks 
diffeomorphisms only spontaneously, while the $\delta$-function-like brane
source in (\ref{RSaction}) breaks diffeomorphisms {\em explicitly}. We 
therefore have a discontinuity between the two setups. That is, for any 
arbitrarily large but finite value of the parameter $\zeta$ gravity in the
above domain wall background is qualitatively different from that 
in the Randall-Sundrum model. In particular, in the former case we always
have an extra equation (\ref{EOM4oGBsmXYY}), which ensures that $\rho$
vanishes everywhere, and this, in turn, leads to the requirement that the
localized matter be conformal. 

{}On the other hand, in the Randall-Sundrum model
there is no analog of (\ref{EOM4oGBsmXYY}), and $\rho$ need not vanish.
As explained in \cite{zuraRS}, 
it is precisely this fact that allows a consistent (classical)
coupling between brane matter and bulk gravity in this setup. This is 
precisely due to the fact that the brane in the Randall-Sundrum model breaks
diffeomorphisms explicitly. There is, however, a price one has to pay for 
this. In particular, even though the graviscalar decouples from the brane 
matter in the infra-red, its coupling to the trace of the corresponding 
conserved energy-momentum tensor 
is non-vanishing in the {\em ultra-violet} \cite{zuraRS}. 
As explained in 
\cite{zuraRS}, at the quantum level
this then generically leads to an inconsistency in the coupling 
between brane matter and bulk gravity in the Randall-Sundrum model.

\subsection{Quantum Instability}

{}As we reiterated in the previous subsection, at the quantum level we expect
an inconsistency in the coupling between bulk gravity and brane matter in the
Randall-Sundrum model. Here we would like to point out that a similar 
conclusion holds for gravity in the above type of
smooth domain wall backgrounds as well.

{}Thus, since gravity is localized, generically the localized matter will not
remain conformal at the quantum level. This then implies that generically
we have an inconsistency in the coupling between localized matter and bulk 
gravity at the quantum level. This gives us a hint that localization of gravity
itself might not be stable against quantum corrections. In fact, this is indeed
expected to be the case \cite{COSM,zuraRS,olindo}. In particular, generic
higher curvature terms actually delocalize gravity. Thus,
inclusion of higher derivative terms of, say, the form
\begin{equation}
 \lambda \int d^Dx \sqrt{-G} R^k
\end{equation}
into the bulk action would produce terms of the form \cite{COSM,zuraRS,olindo}
(the hatted quantities are $(D-1)$-dimensional)
\begin{equation}
 \lambda \int d^{D-1}x dy\exp[(D-2k-1)A]\sqrt{-{\widehat G}}{\widehat R}^k~.
\end{equation}
Assuming that $A$ goes to $-\infty$ at $y \rightarrow\pm\infty$, 
for large enough $k$ the factor $\exp[(D-2k-1)A]$ 
diverges, so that at the end of the day gravity is no longer localized.
In fact, for $D=5$ delocalization of gravity takes place already at the 
four-derivative level once we include the $R^2$, $R_{MN}^2$ and $R_{MNRS}^2$
terms with generic coefficients with the only exception being the 
Gauss-Bonnet combination \cite{COSM,zuraRS,olindo,alberto}.

{}A possible way around this difficulty might be that all the higher curvature
terms should come in ``topological'' combinations (corresponding to Euler
invariants such as the Gauss-Bonnet term) 
so that their presence does not
modify the $(D-1)$-dimensional propagator for the bulk graviton modes
\cite{COSM,zuraRS,olindo,alberto}. That is,
even though such terms are multiplied by diverging powers of the warp factor,
they are still harmless. One could attempt to justify the fact that higher
curvature bulk terms must arise only in such combinations by the fact that
otherwise the bulk theory would be inconsistent to begin with due to the
presence of ghosts. However, it is not completely obvious whether it is
necessary to have only such combinations to preserve unitarity. Thus, in
a non-local theory such as string theory unitarity might be preserved,
even though at each higher derivative order there are non-unitary terms, due
to a non-trivial cancellation between an infinite tower of such terms.

{}Recently, however, a novel approach to this problem
has been proposed in \cite{alberto}. The setup of \cite{alberto} is the
Einstein-Hilbert-Gauss-Bonnet gravity with negative cosmological term.
As was shown in \cite{alberto}, at the special value of the Gauss-Bonnet 
coupling this theory has a codimension one 
{\em solitonic} brane solution. In this solution
the brane is $\delta$-function-like, and gravity is {\em completely} localized
on the brane. That is, there are no propagating degrees of freedom in the bulk,
while on the brane we have purely $(D-1)$-dimensional Einstein gravity. Thus,
albeit the classical background is $D$-dimensional, the quantum theory is
$(D-1)$-dimensional. The aforementioned troubles with delocalization of 
gravity as well as consistency of the coupling between brane matter and bulk 
gravity are therefore absent in the model of \cite{alberto}. In particular,
the brane matter in this model need {\em not} be conformal, and the entire 
setup is stable against quantum corrections\footnote{The only possible 
instability is related to $(D-1)$-dimensional physics, namely, the cosmological
constant on the brane. This solution, however, can be embedded in supergravity
\cite{alberto}, where the solitonic brane is a BPS solution preserving 1/2 
of the original supersymmetries, and the brane cosmological constant 
vanishes.}. 

{}We would like to end our discussion by pointing out that the aforementioned
difficulty with higher curvature terms does not arise in theories with 
infinite-volume non-compact extra dimensions 
\cite{GRS,CEH,DGP0,witten,DGP,zura,zuraCBW,DG}. 
However, in such scenarios consistency of the coupling between bulk 
gravity and brane matter might give rise to additional constraints. Thus, 
in some cases the brane world-volume theory must be
conformal \cite{zura}\footnote{In this case gravity is not localized
as the volume of the transverse space is infinite, so the requirement that
the brane matter is conformal need not be violated at the quantum level. 
It would be interesting to understand if there is any relation between 
such setups and \cite{BKV}. Some speculations on this question were recently 
given in \cite{alberto}.}.

\acknowledgments
{}We would like to thank Gregory Gabadadze for
valuable discussions. 
This work was supported in part by the National Science Foundation.
Z.K. would also like to thank Albert and Ribena Yu for financial support.


\begin{references}

\bibitem{early} 
V. Rubakov and M. Shaposhnikov, Phys. Lett. {\bf B125} (1983) 136.

\bibitem{BK}
A. Barnaveli and O. Kancheli, Sov. J. Nucl. Phys. {\bf 52} (1990) 576.

\bibitem{polchi} J. Polchinski, Phys. Rev. Lett. {\bf 75} (1995) 4724.

\bibitem{witt} P. Ho{\u r}ava and E. Witten, Nucl. Phys. {\bf B460} (1996)
506; Nucl. Phys. {\bf B475} (1996) 94;\\
E. Witten, Nucl. Phys. {\bf B471} (1996) 135.

\bibitem{lyk} I. Antoniadis, Phys. Lett. {\bf B246} (1990) 377;\\
J. Lykken, Phys. Rev. {\bf D54} (1996) 3693.

\bibitem{shif} G. Dvali and M. Shifman, Nucl. Phys. {\bf B504} (1997) 127;
Phys. Lett. {\bf B396} (1997) 64.

\bibitem{TeV} N. Arkani-Hamed, S. Dimopoulos and G. Dvali, 
Phys. Lett. {\bf B429} (1998) 263; Phys. Rev. {\bf D59} (1999) 086004.

\bibitem{dienes} K.R. Dienes, E. Dudas and T. Gherghetta, Phys. Lett. 
{\bf B436} (1998) 55; Nucl. Phys. {\bf B537} (1999) 47; hep-ph/9807522;\\
Z. Kakushadze, Nucl. Phys. {\bf B548} (1999) 205; Nucl. Phys.
{\bf B552} (1999) 3;\\
Z. Kakushadze and T.R. Taylor, Nucl. Phys. {\bf B562} (1999) 78.

\bibitem{3gen} Z. Kakushadze, Phys. Lett. {\bf B434} (1998) 269; 
Nucl. Phys. {\bf B535} (1998) 311; Phys. Rev. {\bf D58} (1998) 101901.

\bibitem{anto} I. Antoniadis, N. Arkani-Hamed, S. Dimopoulos and G. Dvali,
Phys. Lett. {\bf B436} (1998) 257.

\bibitem{ST} G. Shiu and S.-H.H. Tye, Phys. Rev. {\bf D58} (1998) 106007.

\bibitem{BW} Z. Kakushadze and S.-H.H. Tye, Nucl. Phys. {\bf B548} (1999) 180;
Phys. Rev. {\bf D58} (1998) 126001.

\bibitem{Gog} M. Gogberashvili, hep-ph/9812296; Europhys. Lett. {\bf 49} 
(2000) 396.

\bibitem{RS} L. Randall and R. Sundrum, Phys. Rev. Lett. {\bf 83} (1999)
3370; Phys. Rev. Lett. {\bf 83} (1999) 4690.

\bibitem{DGP} G. Dvali, G. Gabadadze and M. Porrati, Phys. Lett. {\bf 
B485} (2000) 208.

\bibitem{DG} G. Dvali and G. Gabadadze, hep-th/0008054.

\bibitem{alberto} A. Iglesias and Z. Kakushadze, hep-th/0011111.

\bibitem{Visser} M. Visser, Phys. Lett. {\bf B159} (1985) 22;\\
P. van Nieuwenhuizen and N.P. Warner, Commun. Math. Phys. {\bf 99} (1985)
141.

\bibitem{zuraCBW} Z. Kakushadze, Phys. Lett. {\bf B491} (2000) 317.

\bibitem{COSM} Z. Kakushadze, Nucl. Phys. {\bf B589} (2000) 75.

\bibitem{zuraRS} Z. Kakushadze, Phys. Lett. {\bf B} (in press),
hep-th/0008128.

\bibitem{olindo} O. Corradini and Z. Kakushadze, Phys. Lett. 
{\bf B494} (2000) 302, hep-th/0009022.

\bibitem{GRS} R. Gregory, V.A. Rubakov and S.M. Sibiryakov, 
Phys. Rev. Lett. {\bf 84} (2000) 5928.

\bibitem{CEH} C. Csaki, J. Erlich and T.J. Hollowood, Phys. Rev. Lett. {\bf
84} (2000) 5932.

\bibitem{DGP0} G. Dvali, G. Gabadadze and M. Porrati, Phys. Lett. {\bf B484}
(2000) 112; Phys. Lett. {\bf B484} (2000) 129.

\bibitem{witten} E. Witten, hep-ph/0002297.

\bibitem{zura} Z. Kakushadze, Phys. Lett. {\bf B488} (2000) 402;
Phys. Lett. {\bf B489} (2000) 207;
Mod. Phys. Lett. {\bf A15} (2000) 1879, hep-th/0009199.

\bibitem{BKV} 
M. Bershadsky, Z. Kakushadze and C. Vafa, Nucl. Phys. {\bf B523} 
(1998) 59;\\
Z. Kakushadze, Nucl. Phys. {\bf B529} (1998) 157;  Phys. Rev. {\bf D58} 
(1998) 106003; Phys. Rev. {\bf D59} (1999) 045007; Nucl. Phys. {\bf B544} 
(1999) 265.



\end{references}
\end{document}